\documentclass[12pt]{article}
\usepackage{amsmath,amssymb,graphicx,mathrsfs}
\usepackage{hyperref}
\newcommand{\be}{\begin{equation}}
\newcommand{\ee}{\end{equation}}
\newcommand{\bea}{\begin{eqnarray}}
\newcommand{\eea}{\end{eqnarray}}

\def\({\left(} \def\){\right)}

\begin{document}

\title{ The sound damping constant \\
for generalized theories of gravity}
 \author{ Ram Brustein \\   Department of Physics, Ben-Gurion
University,\\
    Beer-Sheva, 84105 Israel, E-mail: ramyb@bgu.ac.il \\ \\
A.J.M.  Medved \\ Physics Department, University of Seoul, \\
Seoul 130-743 Korea,
    E-mail: allan@physics.uos.ac.kr}
\date{}
\maketitle


\abstract{

The near-horizon metric for a black brane in Anti-de Sitter (AdS)
space and the metric near the AdS boundary both exhibit hydrodynamic
behavior. We demonstrate the equivalence of this pair of
hydrodynamic systems for the sound mode of a conformal theory.
This is first established for Einstein's gravity, but we then show
how the sound damping constant will be modified, from its Einstein
form, for a generalized theory. The modified damping constant is
expressible as the ratio of a pair of gravitational couplings that
are indicative of the sound-channel class of gravitons. This ratio
of couplings differs from both that of the shear diffusion coefficient
and the shear viscosity to  entropy ratio. Our analysis is mostly
limited to conformal theories but suggestions are made as to how
this restriction might eventually be lifted.





\section{Introduction}
\label{intro}
The near-horizon geometry of  a black brane in an
Anti-de Sitter (AdS) spacetime provides a translationally invariant
and thermally equilibrated background; two of the
characteristic
features of any hydrodynamic theory. Indeed, the long-wavelength
fluctuations of the near-horizon metric are known to
satisfy equations of motion that are completely
analogous to the hydrodynamic equations of a viscous fluid
\cite{hydroI}. The very same statements can be made about the metric
near the AdS boundary. It is, however, quite remarkable  that
 this pair of  effective theories appears to be described
by an equivalently defined set of hydrodynamic parameters
\cite{hydroI,alex-liu,hydroIII,starinets-M};
this, in  spite of their obvious lack of proximity.

The relevant thermodynamic and hydrodynamic parameters --- such as
the entropy density or the various transport coefficients ---
are intrinsic properties of the black brane horizon. Consequently,
such parameters should  be determined by the near-horizon metric. This makes it all the more
phenomenal that the very same parameters are employed
in the AdS boundary theory.
In the context of the gauge--gravity duality,
this  apparent non-locality becomes much more sensible.  The duality relates the AdS boundary hydrodynamics to
the hydrodynamics of strongly coupled gauge theories
\cite{PSS1,hydroIII}. Meanwhile, the dual gauge theory
is supposed to have its thermal properties  ascribed to it,
holographically, by the thermodynamic nature of the black brane.

Most calculations in this genre take place at the outer boundary, as this is the
most convenient surface for relating  the bulk gravitational theory
to its gauge-theory dual. In many ways, however, the most natural setting
is at the black brane horizon, where the various hydrodynamic parameters
are actually defined. The graviton hydrodynamic ``fluid"
can be interpreted as ``living" on the stretched
horizon, and so it would be rather disturbing
if the actual calculations could only be performed on a surface
that is displaced a spacetime away.  Our results will make it clear
that there is really nothing  particularly special about either
the stretched horizon or the  AdS outer boundary.
Rather, all calculations might as well be done on any radial shell
that is external to the horizon.

Strongly coupled gauge theories
provide an intriguing  theoretical laboratory  to
investigate the field of relativistic hydrodynamics.
It is hoped that, by applying the duality to
certain calculations on the gravity side, one
would be able explain the
experimental results of --- for instance ---
heavy-ion collisions \cite{ion}.
However, advancements in this direction has
been somewhat impeded for the following reason:
Studies on the hydrodynamics of the AdS boundary have, for
the most part, been  limited to  Einstein's theory of gravity, which
--- from the gauge-theory perspective ---
corresponds to infinitely strong 't Hooft coupling.
Insofar as the objective is to apply what can be learnt from
the duality  to physically real systems,
one actually requires knowledge about gauge theories at {\em finite}
values of 't Hooft  coupling.

As it so happens,
a strong-coupling expansion on the gauge-theory
side corresponds to
an expansion in the number of derivatives on the gravity
side. Since Einstein's gravity is only a two-derivative theory,
it should be clear that describing a finitely coupled gauge theory
necessitates some sort of extension of Einstein's theory.
To put it another way, any
discernible progress will depend upon our understanding of
the boundary hydrodynamics for theories of {\em generalized} gravity.

In a previous paper \cite{shear}, we were able to establish
two relevant points. First, with the focus  on the
shear channel of fluctuating gravitational modes,
it was shown that
the AdS boundary hydrodynamics
can be translated to and localized on
any radial shell in the accessible spacetime; including
at the (stretched)
horizon of the black brane.
Then, by following \cite{us}, we explicitly demonstrated
how this formalism can
be extended to any generalized (or Einstein-corrected)
gravitational theory.

In \cite{us,shear}, we  used insight from
\cite{BGH} to make a pertinent observation: Various hydrodynamic
parameters of an AdS brane theory can be identified with the
different components of
a (generally)  polarization-dependent gravitational coupling
$\kappa_{\mu\nu}$. Meaning that, for a generic theory, differently
polarized gravitons will effectively have differing  Newton's
constants. As shown in \cite{BGH}, this distinction can be quantified
at a remarkably
rigorous level.  With this prescription, the shear viscosity
to entropy density ratio
$\eta/s$ is generalized from its ``standard" (Einstein)
value of $1/4\pi$ according to \cite{us}
$
\frac{\eta}{s}=\frac{1}{4\pi}\ \frac{ \left(\kappa_{rt}\right)^2}
{\left(\kappa_{xy}\right)^2}\;
$,
with the precise meaning of the subscripts  to be clarified below.
Moreover, the central finding of \cite{shear} was that
the shear diffusion coefficient $D$ is  modified from
its usual expression $1/4\pi T$  into the form
$D= \frac{\kappa^2_{zx}}{\kappa^2_{tx}}\frac{1}{4\pi T}\;$,
with $T$ being the temperature.

One should take note that the coupling ratios for $\eta/s$ and $D$
involve different polarization directions. This is  a natural
consequence of the class of gravitons that is implicated by each of
the  hydrodynamic parameters. In the so-called radial gauge,
the non-vanishing gravitons separate into three decoupled classes
or ``channels":
scalar, shear and sound \cite{PSS2}.
The shear viscosity $\eta$ is most directly associated
with the first of these classes, whereas  the shear diffusion
coefficient $D$ is a characteristic of the second.
As for the third class, one would
analogously associate with it the sound damping
constant $\Gamma$, as well as  the
sound velocity (squared) $c^2_s$.

The purpose of the current paper is to analyze the case of 
sound-mode fluctuations.
A straightforward extension of previous analyses is inhibited by
two technical issues that are intrinsic to  almost any 
rigorous study  of the sound channel.
First, for a non-conformal gauge theory, the sound-channel analysis is highly model
specific.  Second, the same non-conformality induces would-be radial invariants
to vary  with radial position in the bulk. (See \cite{liu} for a discussion.)
As a consequence,  $\Gamma$ and $c^2_s$ are, even for Einstein's gravity,
model-dependent parameters that vary with radial position
in a model-specific way.

We can still be  quite definitive by
restricting  the immediate considerations to  conformal theories.
When conformality is protected by effectively ``switching  off" all massive fields,
the above complications will no longer be of issue.
At the same time, we will still be able to make statements about how
deviations from conformality should influence
the ensuing results. In this sense, the current study can be viewed
as a  significant first step  towards  a fully generic analysis.

Similarly to \cite{shear},
we will begin here by establishing  a direct
connection between sound-mode (conformal) hydrodynamics on the
AdS outer boundary and  on any other radial shell up
to the horizon of the  black brane.
This will be accomplished
by examining the correlator of an appropriately defined
graviton and  verifying that its pole
structure, which determines the associated dispersion relation,
is a radial invariant.

Next, we will determine how this correlator pole is explicitly modified
for a generalized (although still conformal) theory of gravity. This will
enable us to extract the Einstein-corrected form of the
damping constant $\Gamma$. Additionally, we will confirm
that the sound velocity $c^2_s$  remains fixed
at its conformal value.
As also discussed, the very same outcomes
can be deduced through an inspection of
the conservation equation for the dissipative stress tensor.

The paper will conclude with a preliminary discussion of possible extensions
of our analysis to the non-conformal case.

Note that,
to avoid needless repetition, some salient points that are already covered
thoroughly in \cite{shear} (also see \cite{us}) will only be glossed over here.

\section{Sound mode conformal hydrodynamics for Einstein's gravity}

Let us first introduce some
notation and conventions,
as well as establish the basic framework. We will be considering a
black $p$--brane in a $d+1$-dimensional (asymptotically) AdS spacetime.
(Note that $d = p+1 \ge 4$.)
Given translational invariance  and spatial isotropy
on the brane  along with a static
spacetime, the associated metric can
always be expressed in the generic brane form
\begin{equation}
ds^2=-g_{tt}(r) dt^2
+g_{rr}(r) dr^2+ g_{xx}(r)\left(\sum_{i=1}^p
dx_i^2\right)\;,
\label{100}
\end{equation}
where $g_{tt}(r)$ has a simple zero and $g_{rr}(r)$ has a simple pole  at the horizon
$r=r_h$, while $g_{xx}(r_h)$ is finite and positive.
For any $r>r_h$,  these  metric components
are all well-defined and strictly positive functions that  go
asymptotically to their respective AdS values ($L^2/r^2$ for $g_{rr}$,
otherwise $r^2/L^2$)
as $r\rightarrow\infty$.
($L$ is the AdS radius of curvature.)

If the background theory is  conformal, then one
can be much more explicit. Assuming, for the sake
of simplicity, that the brane is electromagnetically neutral,
we obtain the Schwarzschild-like form such that
$g_{xx}=r^2/L^2$ and
$g_{tt}=1/g_{rr}= g_{xx} f(r)$, with
 $f(r)=1-\left( r_h/r \right)^{p+1}$. It is often convenient to
re-express this conformal metric by changing the radial coordinate
to $u=r_h^2/r^2$; then
\begin{equation}
ds^2=-\frac{r_h^2}{L^2 u}f(u)dt^2
+\frac{L^2}{4u^2}\frac{du^2}{f(u)}+\frac{r_h^2}{L^2u}\left(\sum_{i=1}^p
dx_i^2\right)
\;,
\label{2}
\end{equation}
with $f(u)=1-u^{\frac{p+1}{2}}$
and the horizon (outer boundary) now located at $u=1$
($u=0$). When non-conformal theories are discussed,
$u$ will refer to a radial coordinate
that is appropriately defined so as to extend over the same range
of values.

Brane hydrodynamics entails expanding the metric: $g_{\mu\nu}
\rightarrow {\overline g}_{\mu\nu} +h_{\mu\nu}$, with
$h_{\mu\nu}$ representing the fluctuations or gravitons.
Let us  ---  without loss of generality --- specify $x_p$
to be the direction of graviton propagation
on the brane and re-label it as $z$.
It follows that $h_{\mu\nu}\sim {\rm exp}[-i\Omega t +i Q
z]$ (and, otherwise, depending only on $u$),  where
$(\Omega,0,...,0,Q)$ is the $p+1$--momentum of the graviton.

The choice of  radial gauge,
$h_{u\alpha}=0$ for any $\alpha$,
is known to separate the non-vanishing fluctuations into
three decoupled classes
\cite{PSS2}.
Our class of  current interest --- namely, the sound channel ---
includes the non-vanishing diagonal gravitons $h_{\alpha\alpha}$
($\alpha\ne u$) along with $h_{tz}$.

Let us take note of the sound-mode dispersion relation
$\Omega=\pm c_s Q - i\Gamma Q^2 +{\cal O}(Q^3)\;$
or, equivalently (given that the hydrodynamic
or long-wavelength limit  is in effect),
\begin{equation}
\Omega^2 =c^2_s Q^2 +i2\Gamma\Omega Q^2 +{\cal O}(Q^4)\;.
\label{200}
\end{equation}
Here, $c_s^2$ is the sound velocity (squared) and $\Gamma$
is the sound damping constant. For a $p+1$-dimensional conformal  theory,
$c_s^2=1/p$ and  $\Gamma$
is directly proportional to the shear viscosity to entropy
density ratio times the inverse temperature:
$\Gamma=\frac{p-1}{p}\frac{1}{T}\frac{\eta}{s}$.
So that,
for a $p$-brane theory of Einstein's gravity, one can deduce
that $\Gamma=\frac{p-1}{p}\frac{1}{4\pi T}$ \cite{sound,sound-dis},
where $T$ is the coordinate-invariant Hawking temperature of the brane.
For this conformal case, $T=(p+1)r_h/4\pi L^2$.
Meanwhile, for a non-conformal theory, both parameters
can differ appreciably from their conformal values. For $\Gamma$,
this  model-specific deviation is expressible in terms of the bulk
viscosity $\zeta$.

For a complete derivation of the sound-mode correlator (which is not needed here),
one can follow the by-now standard prescription  as  documented
in, for instance, \cite{dam,pavel,mas}.
The first step is to identify a gauge-invariant combination
of the sound-mode fluctuations $H_{tt}=(1/g_{tt}) h_{tt}$, $H_
{zz}=(1/g_{xx}) h_{zz}$, $H_{tz}=(1/g_{xx}) h_{tz}$ and
$H_{X}=(1/g_{xx})\frac{1}{p-1}\left(\sum_{i=1}^{p-1}
h_{x_ix_i}\right)$:
\begin{equation}
Z=q^2 \frac{g_{tt}}{g_{xx}} H_{tt} +2q\omega H_{tz}
+ \omega^2 \left[H_{zz}-H_X\right]
+q^2 \frac{{g_{tt}}^{\prime}}{{g_{xx}}^{\prime}} H_{X}\;.
\label{3}
\end{equation}
Here,  a prime indicates a differentiation
with respect to $u$; while
$\omega=\Omega/2\pi T$ and $q=Q/2\pi T$ represent, respectively,
a dimensionless frequency and wavenumber.
In the hydrodynamic limit,  $\omega$ and $q$ are both vanishing although not {\em a priori} at
the same rate.

The conformal version of the solution for $Z$ can readily be
extracted out of the existent literature --- for instance, \cite{pavel,mas,fujita}.
For the appropriately chosen boundary conditions (as discussed below), one finds that
\begin{equation}
Z = C f(u)^{-\frac{i\omega}{2}}\left[Y(u)-\frac{\omega^2}{q^2}p
-i\omega (p-1)f(u) +{\cal O}(q^2,\omega^2)\right]\;,
\label{5}
\end{equation}
where $C$ is an integration constant (to be fixed by
 normalization considerations) and
 we have defined $Y(u)\equiv {g_{tt}}^{\prime}/{g_{xx}}^{\prime}$.
For this theory in particular, $Y=(f/u)^{\prime}/(1/u)^{\prime}=f-uf^{\prime}$,
which is everywhere positive, non-vanishing and ${\cal O}(1)$.
Also note that $Y=1$ on the outer boundary.

A specified pair of boundary conditions determines the solution for $Z$. At the horizon $u=1$,
the solution should be that of an incoming plane wave, which determined the form of Eq.~(\ref{5}).
In addition, the  so-called Dirichlet boundary condition still needs to be imposed.
It has become almost traditional to
single out the AdS boundary and choose $u^*=0$ as
the radius at which this condition is enforced; however, one can freely
impose this condition at any fixed radius $u^*$ within $0 \le u^* < 1$.
The Dirichlet boundary condition necessitates
that $Z(u)$ is, {\it prior} to its normalization (see below), vanishing as $u\rightarrow u^*$.
Applying this condition to Eq.~(\ref{5}), we
promptly obtain  the associated dispersion relation
\begin{equation}
\omega^2= q^2 \frac{1}{p}Y(u^*) +i\omega q^2 \frac{p-1}{p} f(u^*) +{\cal O}(q^4)\;,
\label{500}
\end{equation}
where it is now clear that $q$ and $\omega$
are of the same order in the hydrodynamic limit.
Let us choose, for instance, the ``orthodox" boundary location
of $u^*=0$, so that  Eq.~({\ref{500}) leads
to $\omega^2 =q^2/p +i\omega q^2 (p-1)/p +{\cal O}(q^4)$.
Comparing this  to
the standard  dispersion relation in Eq.~(\ref{200}),
one can readily verify the expected identifications
$c^2_s=1/p$ and $\Gamma=(p-1)/(4\pi p T)$ for a
conformal theory.

One further normalization condition  that complements the Dirichlet boundary
condition is that $Z$, rather than vanishing at
$u^*$, should ultimately be normalized to unity there. This can be achieved by the unique choice
\begin{equation}
C^{-1}=  \left[Y(u^*)-\frac{\omega^2}{q^2}p
-i\omega (p-1)f(u^*) \right]\;.
\end{equation}
Let us take notice that, given the associated dispersion relation,
$C^{-1}$ is a vanishing quantity as it must be to obtain
a finite value of $Z(u^*)$.

The normalized  value of the field mode is simply $1$,  and so one might wonder as
to the physical significance of the implied dispersion relation.
However, we are simply  using the standard ``trick" of field-theoretic calculations  
to obtain the  pole in the correlator.
The properly normalized correlator $G_{ZZ}$ for this gauge-invariant
variable  (up to an inconsequential numerical
factor) is given by the boundary residue of the canonical
term in the bulk action or $G_{ZZ}\sim ZZ^{\prime}\vert_{u=u^*}$.
It should not be difficult to convince oneself that, at the
leading hydrodynamic order, this quantity goes as $G_{ZZ}\sim C \cdot{\cal O}(q^0)$,
which is notably divergent and of  finite hydrodynamic order.
A proper accounting of the metrical factors in the action
--- namely, the product $\sqrt{-g}g^{uu}$ --- reveals that
there are no other hidden zeros or infinities in this
calculation at any permissible value of $u^*$.

What is really significant here is that ---
from the quasinormal-mode perspective of brane hydrodynamics \cite{pavel} ---
the pole in the correlator assigns a
clear physical credence to  Eq.~(\ref{500}) as the spectrum
for the dissipative modes of the black brane.
However, one could (and should!) be rightfully
concerned that this dispersion relation appears
to vary as the Dirichlet-boundary surface is moved radially
through the spacetime. This is not only in conflict with
intuitive expectations but with the analysis of \cite{liu},
where it is made evident that (inasmuch as the theory is conformal)
both the sound mode and its correlator should be radial invariants.

We can readily account for the undesirable factor of $f(u^*)$
in the second term of Eq.~(\ref{500}): As
detailed in \cite{shear},
$\omega$ and $q$ should naturally be sensitive to the
the effects of a gravitational redshift. It was then
argued --- in the context of shear modes --- that
consistency of the hydrodynamic expansion along with
protection of the incoming boundary condition necessitates
that $\omega$ remains fixed while $q^2$ scales as $f^{-1}$.
That is, $\omega(u)=\omega_{b}$
and $q^2(u)=q^2_b/f(u)$ (with the subscript $b$ indicating
the outer-boundary value of a quantity).
Then, since  the gravitational redshift should not be able to
discriminate between the different channels  being probed, it
follows that these same relations should persist for the
current case.

This brings us to the first term in Eq.~(\ref{500}),
which has the awkward appearance of $Y(u^*)$ to be dealt with.
Clearly, this will require new inputs. The key here
is the association of this term with $c^2_s$;
{\it cf}, Eq.~(\ref{200}). Normally, the sound velocity
of a hydrodynamic fluid is presented as the variation of the
pressure with respect
to the energy density or $c_s^2={\delta P}/{\delta\epsilon}$.
This cannot, however, be a universally accurate account.
A closer  look at the derivation of the sound dispersion
relation (see, {\it e.g.}, \cite{sound-dis}) reveals that the actual variation
which enters under the guise of the sound velocity comes packaged
in the term
$\left({\delta T^{zz}}/{\delta T^{tt}}\right)\partial_z^2 T^{tt}$,
where $T^{\alpha\beta}$
is the stress tensor for the brane theory. For a flat or an effectively flat brane,
such as at the AdS outer boundary, this
distinction is of no
consequence, but this is not a general truism. On a ``warped" brane,
rather,
$T^{zz}=g^{zz}P$ and $T^{tt}=-g^{tt}\epsilon$.
Hence, the correct statement about the relevant term is (by way
of the chain rule)
\begin{equation}
\frac{\delta T^{zz}}{\delta T^{tt}}\partial_z^2  T^{tt} =
g^{zz}\frac{\delta P}{\delta \epsilon} \partial_z^2 \epsilon +
g^{tt}\frac{P}{\epsilon}\frac{\partial_u g^{zz}}{\partial_u g^{tt}}\partial_z^2 \epsilon \;.
\label{700}
\end{equation}

We will now argue that the first term on the right-hand side of Eq.~(\ref{700})
is parametrically smaller than the second and, thus,
the former can be disregarded for current purposes. It follows from
the thermodynamic relation $sT=\epsilon +P$ and the infinite transverse volume of the brane that
 ${\partial P}/{\partial\epsilon}=0$.
So, to the leading non-vanishing order,
$\delta P= \frac{1}{2}
\frac{\partial^2 P}{\partial \epsilon^2} \left(\delta\epsilon\right)^2$
or $\frac{\delta P}{\delta \epsilon}\sim \frac{\delta\epsilon}{\epsilon}\ll 1\;$.
On the other hand, ${P}/{\epsilon}=\frac{1}{p}$ is of the order of unity.
Now, comparing  $g^{tt}\frac{\partial_u g^{zz}}{\partial_u g^{tt}}$ with $g^{zz}$,
one will find that the ratio of these
quantities is of ${\cal O}(1)$.

Having deemed the first term in Eq.~(\ref{700}) as inconsequential,
we need  only to evaluate the second.
Since the brane metric is diagonal (so that
$g^{\alpha\alpha}=g^{-1}_{\alpha\alpha}$), the right-hand side reduces to
\begin{equation}
\frac{\delta T^{zz}}{\delta T^{tt}}\partial_z^2 T^{tt}=
\frac{P}{\epsilon}\frac{g_{tt}}{g_{xx}^2}Y^{-1}\partial_z^2\epsilon+\cdots\;,
\label{1000}
\end{equation}
where we have returned to the brane notation of Eq.~(\ref{100})
(so that $g_{tt} >0$) and recalled
the definition of $Y(u)$ beneath
Eq.~(\ref{5}).

Actually, all other terms in
the dispersion relation contain a spatial component of the
brane stress tensor, and so share a common factor of $g_{xx}^{-1}$.
Hence, we can strip off one factor of this from the right side of Eq.~(\ref{1000}). Next, let us
identify $P/\epsilon$ as the sound velocity as measured
on the outer AdS boundary and
everything  to the left of $\partial_z^2\epsilon$  as the
sound velocity as measured on a radial shell of arbitrary
radius. Then it follows that the sound velocity scales
relative to the outer boundary
as
\begin{equation}
\left[c^2_s\right]_u = Y^{-1}(u)\frac{g_{tt}(u)}{g_{xx}(u)} [c^2_s]_b\;,
\end{equation}
where a subscript of $u$ denotes the value of a parameter at that radius.
Calling again on our conformal-theory notation, let us take note that, by definition,
$f=g_{tt}/g_{xx}$, and so the sound velocity equivalently
scales as $f/Y$.

Next, let us re-express Eq.~(\ref{500}) in a way that makes the scaling
properties of the parameters explicit:
\begin{equation}
\omega_{u^*}^2= q_{u^*}^2 Y(u^*) [c^2_s]_{u^*}
+i\omega_{u^*} q_{u^*}^2 \frac{p-1}{p} f(u^*) +{\cal O}(q_{u^*}^4)\;.
\label{2000}
\end{equation}
Here, we have made the identification $\frac{1}{p} \rightarrow [c_s^2]_{u^*}$ on the basis
that the Dirichlet-boundary surface  is where
the sound velocity should be calibrated to
its conformal value --- just like it is the Dirichlet surface
that defines where the field $Z$  is exactly unity. We can now apply the
previously discussed scalings ($q^2\sim 1/f$, $c^2_s\sim f/Y$ and an invariant $\omega$)
to convert the above expression into one that involves only the outer-boundary  values of
the parameters. Also recalling that $f=Y=1$ at the AdS boundary $u=0$, we then have 
\begin{equation}
\omega_{b}^2= q_{b}^2 Y(0) [c^2_s]_{b} +i\omega_{b} q_{b}^2 \frac{p-1}{p} f(0)  +{\cal O}(q_b^4)\;.
\label{3000}
\end{equation}
But  this is precisely what would have  been obtained  had we made
the choice of $u^*=0$ in the first place.  Hence, the dispersion relation
is indeed a radial invariant and, by direct implication, the correlator is
as well.

\section{Sound mode conformal hydrodynamics for generalized theories of gravity}

Next on the agenda, we will investigate as to how
the scenario changes when the theory is extended from Einstein's gravity. 
It will be shown that, for a quite general
(although still conformal) gravity theory, the damping constant
is modified  in a very precise way. Meanwhile, the
sound velocity is shown to be unmodified, as must
be the case for a conformal theory.
These tasks will be accomplished by
examining the (modified) pole of the
just-discussed correlator. These generalizations
will be further supported by a
simple argument that is based upon inspecting
the conservation equation  that gives rise
to the sound dispersion relation.

By a generalized gravity theory, we have in mind
a Lagrangian that can be expressed as Einstein's form
plus higher-derivatives terms.
If Einstein's gravity is ``non-trivially" modified
by these corrections --- meaning that the general Lagrangian
can {\em not} be  converted into Einstein's form  by a  field
redefinition ---
then the gravitational coupling
is no longer as simple as $\kappa^2_E={\rm constant}$.   Rather,
the coupling (or effective Newton's constant)
can be expected to depend on the
polarization of the gravitons being probed. We will
denote this dependence by expressing the general couplings
as $\kappa_{\mu\nu}$.

It is now well understood as to how one should calculate these couplings
for a given theory \cite{BGH,us,shear}. These formalities need not concern
the present discussion, although a schematic understanding of how the couplings come about
should prove useful. One begins by writing the Lagrangian
as a perturbative expansion in powers of the metric fluctuations or $h$'s.
Of particular significance are the terms that are quadratic in $h$
and contain exactly two derivatives. For such terms,
the gravitational couplings are identified on the premise
that $h_{\mu\nu}\rightarrow \kappa_{\mu\nu}h_{\mu\nu}$
leads to a {\em canonical} kinetic term for the $\mu\nu$-polarized graviton.

As it turns out, the gravitational couplings
are expressible strictly in terms of the metric at the horizon.
Like the metric, they are typically radial functions; however,
at the level of a two-derivative
expansion of the Lagrangian, the
couplings
can safely  be treated as (polarization-dependent)
constants. Moreover, since the horizon is the true arena for black brane
hydrodynamics, this locality is quite natural
and  falls in line with other
parameters, such as the entropy and shear viscosity,
being intrinsic properties of this special surface.

Let us re-emphasize that any given hydrodynamic parameter should be modified
according to the class of gravitons that it probes.
By working in the radial gauge and  then restricting to the
decoupled set of  modes that defines the sound channel,
we are limited to a select class. Namely, the $zz$, $tt$, and $tz$-polarized
gravitons, as well as the ``trace mode",
which can be identified with $H_X$ in Eq.~(\ref{3}).

When the theory is conformal, we can anticipate a further limitation.
To elaborate, in obtaining  the solution for $Z$
(see, {\it e.g.}, \cite{pavel,mas,fujita}), one finds
that the $H_{tt}$ mode  makes no direct contribution to  Eq.~(\ref{5}).
(This is not at all true when conformality is broken.)
Recalling that the gauge--gravity duality identifies
the $tt$-polarized gravitons with fluctuations in the energy density,
we suspect that this null contribution is another manifestation
of the suppression of the  variation     
$\delta P/\delta \epsilon$ (as discussed in the previous section). 
On this basis, it seems reasonable to suggest that the $tt$ fluctuations
can be excluded from a conformal theory  
in  the hydrodynamic limit.

As implied above,
the modifications of interest can be extracted from
the pole structure of the (generalized) correlator $G_{ZZ}$.
Critical to this procedure is
the identification of the gravitational coupling
$h_{\mu\nu}\rightarrow \kappa_{\mu\nu}h_{\mu\nu}$,
which persuades us
to adapt the gauge-invariant variable $Z$ of Eq.~(\ref{3}) as follows:
\begin{equation}
Z= 2q\omega\kappa_{tz} H_{tz} + \omega^2 \kappa_{zz} H_Z
+q^2Y\kappa_{zz}H_X \;,
\label{800}
\end{equation}
where  $H_Z\equiv H_{zz}-H_X$, the non-contributing mode
$H_{tt}$ has been dropped and, as before,
$Y={g_{tt}}^{\prime}/{g_{xx}}^{\prime}$.
Also,  the spatial isotropy of the brane has enabled us
to make the convenient substitution
$\frac{1}{p-1}\sum_{i=1}^{p-1}\kappa_{x_ix_i}\rightarrow \kappa_{zz}$.

The scaling properties of
the damping constant can now be determined with a methodology
akin to dimensional
analysis: {\em First}, redefine the  wavenumber and
the  frequency (and other parameters as necessary)
with a scaling operation, {\em second},  re-express
the solution in terms of these revised parameters and, {\em third},
interpret the modified pole structure.
With regard to  the first step, it
is actually necessary  to fix $\omega$, otherwise the incoming boundary
condition at the horizon would be jeopardized. We are, however,
free at this level of analysis to change
the normalization of $Z$. On this basis, we arrive at
\begin{equation}
Z = 2q\omega\frac{\kappa_{tz}}{\kappa_{zz}}H_{tz}+\omega^2 H_Z
 + q^2YH_X
\end{equation}
or
\begin{equation}
Z = 2\widetilde{q}\omega H_{tz}+\omega^2 H_Z
+\widetilde{q}^2\widetilde{Y}H_X\;,
\end{equation}
with
\begin{eqnarray}
\widetilde {q}&\equiv& q\frac{\kappa_{tz}}{\kappa_{zz}}\;, \cr
\widetilde{Y}&\equiv& Y\frac{\kappa^2_{zz}}{\kappa^2_{tz}}\;.
\end{eqnarray}
By invoking $Y\rightarrow\widetilde{Y}$, we do not mean to suggest that
this function actually gets rescaled. Rather, the presence of $Y$
in Eq.~(\ref{500}) for $Z$ represents a direct contribution from $q^2H_X$,
which --- after rescaling $q^2$ --- picks up
the extra factor $\kappa^2_{zz}/\kappa^2_{tz}$.

Since the couplings can be regarded as constants, the
solution  in Eq.~(\ref{5}) is formally  unchanged and need only be
rewritten in terms of the rescaled parameter. By this logic,
the same can be said about the dispersion relation in
Eq.~(\ref{500}), which takes on the modified form
\begin{equation}
\omega^2=
\widetilde{q}^{\,2} \frac{1}{p}\widetilde{Y}
 +i\omega \widetilde {q}^{\,2} \frac{p-1}{p}f\frac{\kappa^2_{zz}}{\kappa^2_{tz}}
\; .
\end{equation}
Taking $u^*=0$ and then comparing directly to Eq.~(\ref{200}),
we can promptly extract the
damping constant for a generalized
(but conformal) theory of gravity:
\begin{equation}
\Gamma= \frac{\kappa^2_{zz}}{\kappa^2_{tz}}\frac{p-1}{ p }\frac{1}{4\pi  T}\;
\label{15}
\end{equation}
and, as advertised, the sound velocity is clearly unmodified.

Let us briefly comment upon the significance of this result. It is commonplace,
for a conformal theory,
to relate the sound damping constant directly to the  
shear diffusion coefficient
or (equivalently) the
shear viscosity to entropy ratio: $\Gamma =\frac{p-1}{p} D$ and
$\Gamma = \frac{p-1}{p}\frac{1}{T}\frac{\eta}{s}$ respectively.
This is all indisputably true
for an Einstein theory of gravity; however, as we have now shown,
these relations can not be taken verbatim for a generalized theory.
To be clear, let us compare Eq.~(\ref{15}) to our prior results 
from \cite{shear} $D= \frac{\kappa^2_{xz}}{\kappa^2_{tz}}  
\frac{1}{4 \pi T}$  and from \cite{us}  $\eta/s =\frac{1}{4 \pi} 
\frac{\kappa^2_{rt}}{\kappa^2_{xz}}  $  
(where $x$ and $z$ could be  any pair of orthogonal directions
on the brane and note that, in general, $\kappa_{xz}\neq \kappa_{zz}$).
It should now be evident that both of the above relations
for $\Gamma$ 
will generally be modified
for an Einstein-corrected theory.

The very same outcome as in Eq.~(\ref{15}) can be surmised from the $z$-component
of the conservation equation for the dissipative stress
tensor; with this being the equation that gives rise to
the sound-mode dispersion relation (see, {\it e.g.}, \cite{sound-dis}).
An inspection of this conservation equation
$\partial_t T^{tz} + \partial_z T^{zz} =0$ and
the steps leading up to
the dispersion relation (\ref{200}) is quite revealing.
It is the $tz$ component of the stress
tensor that accounts for the $\Omega^2$ term in
Eq.~(\ref{200}), whereas the $zz$ component gives
rise to  the $Q^2\Omega$ term.
Now, given a gravitational pedigree for the
hydrodynamic modes,
it is natural to associate a coupling of $\kappa_{\mu\nu}^2$
with the $\mu\nu$ component of the stress tensor.
Hence, we anticipate that, for a generalized gravity theory,
the conservation equation
should really be $\kappa^2_{tz}\partial_t T^{tz} +
\kappa^2_{zz}\partial_z T^{zz} =0$.
Similarly,  we can expect
the dispersion relation to take on the modified form
$\Omega^2 =  \frac{\kappa^2_{zz}}{\kappa^2_{tz}}
\left[i2\Gamma\Omega Q^2\right]+\;...\;$.
(with the dots referring to the sound-velocity and
higher-order terms).
Absorbing this  ratio of couplings into the
damping constant,
we have precisely the same generalized
form as obtained in Eq.~(\ref{15}).

Naively, this latter argument would also suggest that $c_s^2$ scales
in the same way as $\Gamma$, given that both are associated
with the same $zz$ component of the stress tensor.
However, this is not really correct: The sound velocity is
associated with the variation of the pressure; with the pressure having
originated from the non-dissipative background part of the stress tensor.
Meanwhile, the other terms in the dispersion relation are strictly
associated with the fluctuations or leading-order dissipative part.
On this basis, we would not anticipate the sound velocity to be scaled
for a generalized (conformal) theory; again in compliance
with the previous analysis.

\section{Discussion: Some aspects of the non-conformal case}

To summarize, we have demonstrated two
important outcomes for the sound-mode conformal hydrodynamics
of an AdS brane theory. First, we have confirmed, for
Einstein's theory, that the hydrodynamics at the outer
boundary is equivalent to that of any other radial
shell up to (and including at) the stretched horizon.
 Second, we have shown
--- quite precisely --- how the sound velocity and damping coefficient
 will be  modified for a generalized (but conformal) theory
of gravity. More specifically, $c^2_s$ is unaffected, whereas
$\Gamma$ is scaled by a particular ratio of (generalized) gravitational couplings.
 Further note that, inasmuch as the couplings can
be treated as constants, the former outcome will  carry through
unfettered for any Einstein-corrected conformal theory. 

It is also of some interest to reflect upon how a non-conformal
theory would impact upon our findings. Let us first consider the 
issue of radial invariance for Einstein's theory.
Clearly, this
invariance for the correlator depended,
in large part, on being able to
disregard the first
term in Eq.~(\ref{700}). However,
the introduction of a massive field into the spacetime
(a prerequisite for breaking conformality)
would
be tantamount to the inclusion of  a chemical potential into the thermodynamics.
Such an inclusion would then negate our previous argument
for the suppression of the scrutinized term; in particular,
$sT = \epsilon +P$ could no longer be true.
Hence,
there could no longer be any reason to expect
that $\frac{\delta P}{\delta \epsilon}$ is a parametrically
small quantity for a non-conformal theory  --- meaning that  the 
radial scaling of the sound
velocity would certainly be more complicated. However, that this
deviation from  the conformal calculation is seemingly
encapsulated in the single variation $\frac{\delta P}{\delta \epsilon}$
gives one hope of being able to
describe even
the fully general situation by way of a radial ``flow" equation.
 Although, it should be kept in mind that a
further breach of radial invariance is possible (if not probable)
from additional terms that would (almost inevitably) appear
in the ${\cal O}(q^1)$  solution for $Z$.

For the case of generalized gravity, the state of affairs can become significantly  more
convoluted for a non-conformal theory. Here, the first order of business is
to re-incorporate the previously disregarded $tt$ mode --- but then what?
Well, at a first glance, the situation does not appear to look too bad. For the reason discussed
at the end of the prior section, we would not expect the sound velocity to
be modified irrespective of the generalized gravitational couplings.
As for the damping coefficient, one can show that $H_{tt}$ makes no contribution
to this particular term, so it seems reasonable to suggest that $\Gamma$
maintains its modified form of Eq.~(\ref{15}).

It is, however, a nearly certain likelihood that
the situation can not be  as simple
as so far discussed. For a non-conformal theory, there
is an inevitable mixing between $H_X$ and the massive bulk
fields, and it is not yet clear as to how
this mixing might effect the scaling relations
for either $\Gamma$ or $c_s^2$ (with both of these
being directly implicated with the ``polluted" $H_X$ mode).
Certainly, a mode formed out of $H_X$ and
some, for instance, massive scalar field, could no longer
have an effective coupling as trivial as $\kappa_{zz}$.

The main issue of non-conformal treatments is that,
due to the high degree of model dependence
in the formalism,
very little can be said in a generic sense.
There has, however, been some recent progress in
such a direction \cite{gubser,springer}. These
papers indicate that a better starting point might
be to look at certain classes of non-conformal
theories, as opposed to the ``extreme limiting cases"
of a specific model or completely generality.
Work along this line is only at a preliminary
stage.

\section*{}
{\bf Acknowledgments:}
The research of RB was supported by The Israel Science Foundation
grant no 470/06.
The research of AJMM is supported by the University of Seoul.

\end{document}